\documentclass[12pt]{article}
\usepackage{amssymb}
\usepackage{amsfonts}
\usepackage{amsmath}
\usepackage{mathtools}  
\usepackage{bm}
\usepackage{latexsym}
\usepackage{epsfig}
\usepackage{bbm}
\usepackage{enumerate}
\usepackage{float}
\usepackage{color}
\usepackage{gensymb}
\usepackage{siunitx}
\usepackage{color}
\usepackage{hyperref}
\usepackage[round,authoryear]{natbib}
\usepackage[T1]{fontenc}
\usepackage[utf8]{inputenc}
\usepackage{authblk}
\usepackage{appendix}
\usepackage{blindtext}
\usepackage{lineno}
\usepackage{nth}
\usepackage{bbm}
\usepackage{setspace}

\usepackage[
        a4paper,
        left=2.5cm,
        right=2.5cm,
        top=3cm,
        bottom=3cm]{geometry}
\begin{document}

\title{New Exploratory Tools for Extremal Dependence:\\ $\chi$ Networks and Annual Extremal Networks}

\author{Whitney K. Huang\footnote{Pacific Climate Impacts Consortium, University of Victoria. E-mail: \href{whuang@samsi.info}{\nolinkurl{whuang@samsi.info}}}, Daniel S.  Cooley\footnote{Department of Statistics, Colorado State University. E-mail: \href{cooleyd@stat.colostate.edu}{\nolinkurl{cooleyd@stat.colostate.edu}}}, Imme Ebert-Uphoff\footnote{Department of Electrical and Computer Engineering, Colorado State University. E-mail: \href{iebert@colostate.edu}{\nolinkurl{iebert@colostate.edu}}}, Chen Chen\footnote{Department of the Geophysical Sciences and Statistics, University of Chicago. email: \href{chenchen1@uchicago.edu}{\nolinkurl{chenchen1@uchicago.edu}}}, Snigdhansu Chatterjee\footnote{School of Statistics, University of Minnesota. email: \href{chatt019@umn.edu}{\nolinkurl{chatt019@umn.edu}}}}

\vspace{2in}

\date{\today}

\maketitle

\begin{abstract} 
Understanding dependence structure among extreme values plays an important role in risk assessment in environmental studies. In this work we propose the $\chi$ network and the annual extremal network for exploring the extremal dependence structure of environmental processes. A $\chi$ network is constructed by connecting pairs whose estimated upper tail dependence coefficient, $\hat \chi$, exceeds a prescribed threshold. We develop an initial $\chi$ network estimator and we use a spatial block bootstrap to assess both the bias and variance of our estimator. We then develop a method to correct the bias of the initial estimator by incorporating the spatial structure in $\chi$. In addition to the $\chi$ network, which assesses spatial extremal dependence over an extended period of time, we further introduce an annual extremal network to explore the year-to-year temporal variation of extremal connections. We illustrate the $\chi$ and the annual extremal networks by analyzing the hurricane season maximum precipitation at the US Gulf Coast and surrounding area. Analysis suggests there exists long distance extremal dependence for precipitation extremes in the study region and the strength of the extremal dependence may depend on some regional scale meteorological conditions, for example, sea surface temperature.

\end{abstract}

\doublespacing
\section{Introduction} \label{sec1}

The 2017 hurricane season was the costliest tropical cyclone season on record \citep{halverson2018}.
The season had 17 named storms, 10 hurricanes, and 6 major hurricanes. Great damage was done to US territories by three major hurricanes, Harvey, Irma, and Maria. These three storms were distinct both in location and timing: Harvey impacted Texas and Louisiana in mid-to-late August, Irma impacted Florida in early September, and Maria devastated Puerto Rico in late September. A standard treatment of the data from these three storms would likely treat them as independent events, due to their temporal and spatial distance. However, it is largely acknowledged that these storms are related, arising from conditions in 2017 conducive to tropical cyclone formation and intensification \citep{emanuel2011,trenberth2018}.
While it is important to assess local impacts of each storm, understanding the connection of hurricane-induced extreme precipitation is also important for insurance companies or government agencies (e.g., Federal Emergency Management Agency (FEMA)) who need to prepare for the possibility of responding to multiple extreme events in a year. Thus we seek to use an approach that studies such phenomena in context of each other, specifically using a spatial network type approach.

The concept of networks has been used in climate science for about 15 years. The connection was first established by
Tsonis and Roebber \citep{tsonis2004,tsonis2006}, who introduced the concept of {\it climate networks} to study the dynamics of Earth's climate system. A climate network consists of nodes (e.g., grid cells from a climate model output/gridded reanalysis data product) and edges. Two nodes are connected by an edge if the strength of a (pair-wise) dependence measure between the corresponding pairs of time-series is ``strong'' enough. For example, \cite{tsonis2004} connected nodes with correlation greater than 0.5.
This approach has been applied to many climate applications \citep[e.g.][]{tsonis2006,tsonis2008,donges2009,steinhaeuser2011} to gain insights into the dynamics of the climate system over many spatial scales, e.g., teleconnection paths. 

Other important network types in climate science are {\it event synchronization networks} \citep{quiroga2002event,malik2010spatial,malik2012analysis,boers2014south}, where connections are defined based on whether extreme events at one point commonly coincide (within the span of a selected time interval) with extreme events at another point;
{\it phase synchronization networks} \citep{yamasaki2009climate}, where the signals at each point are viewed as oscillations, and connections are defined based on the apparent coupling between those oscillations; and {\it causal networks} \citep{RungeDiss,ebert2014causal,zerenner2014gaussian,kretschmer2016using}, where connections are defined to represent potential cause-effect relationships.
Causal networks can be established by many different methods. For example, they can be based on probabilistic graphical models \citep{Pearl:2000,SpGlSc:2000} (for their use in climate science, see for example \citep{ebert2014causal}); on combining the concept of  Granger causality with vector autoregression (VAR) \citep{lutkepohl2007} or LASSO \citep{tibshirani1996lasso} methods, as known as Graphical Lasso or GLasso, \citep{yuan2007,friedman2008sparse}  (for their use in climate science see for example \citep{YanCI2012}); or on Gaussian models \citep{zerenner2014gaussian}. Note that the majority of these network types are closely related to the concept of correlation. Namely, correlation networks use correlation itself \citep{tsonis2004}; many probabilistic graphical models are derived using partial correlation \citep{ebert2014causal};VAR and GLasso approaches approximate regression coefficients, which can be expressed in terms of partial correlation; and Gaussian models approximate the precision matrix (i.e., inverse covariance matrix), whose elements can be interpreted as partial correlation coefficients. Only two network types mentioned above are not based on correlation, namely phase synchronization networks, which with their assumption of oscillators do not appear to be a good match to model extremes, and event synchronization networks, which are in fact most closely related to the concepts discussed here, but are not based on well understood statistical measures.

Correlation measures the \textit{linear relationship} between a bivariate random vector around the center of its distribution and hence is poorly suited for describing dependence in the tail, i.e., the dependence between extreme events. In contrast, several dependence measures designed specifically for extremes have been proposed in the literature and those are likely to be more appropriate than correlation to construct networks that {\it focus} on analyzing relationships between extremes. 
Three examples are the {\it upper tail dependence}, aka 
\textit{tail dependence coefficient} ($\chi$) \citep{coles1999}, 
the {\it extremal coefficient} ($\theta$) \citep{smith1990,schlather2003}, and the F-madogram ($\nu$)  \citep{cooley2006,naveau2009}. See Chapter 2 of \cite{YanDey:book} for a review. The upper tail dependence is defined as 
\begin{equation} \label{chi}
\chi = \lim_{u \to 1^{-}}\chi(u) =  \lim _{u \to 1^{-}}\mathbb{P}(F_{2}(X_{2}) > u|F_{1}(X_{1}) > u),
\end{equation}
where $X_{1}$ and $X_{2}$ are two random variables and $F_{1}, F_{2}$ their corresponding cumulative distribution function (CDF). Thus the $\chi$ measure is the (limiting) conditional probability of one random variable being extreme, given that the other random variable is extreme. Note that the $\chi$ measure is symmetric in terms of which variable is chosen to be conditioned upon, due to the standardization of the marginal CDFs.

In Section \ref{Sec2}, we introduce a new network type, the $\chi$ network, where we use the aforementioned upper tail dependence ($\chi$) to explore the spatial dependence structure in extremes. We briefly proposed this idea at a workshop (\cite{CI2018_network_paper}), but it is fully developed for the first time here.
Specifically, we construct a $\chi$ network by connecting the pairs of nodes whose $\hat{\chi}\text{'s}$ exceed some specified value (see Sec.~\ref{Sec2.2} for details). However, we find that a straight forward implementation of this procedure does not perform well as it introduces many spurious network connections (see Sec.~\ref{Sec2.1}). This problem, which can be described as a {\it network bias}, is not unique to $\chi$ networks.  In fact, we found that the same bias can occur in traditional correlation networks as well, although, to the best of our knowledge, this issue has never been discussed for correlation networks. In Sec.~\ref{Sec2.4} we describe a method to alleviate this issue by incorporating the spatial structure on $\chi$ observed from the empirical estimates to reconstruct the $\chi$ network. Although we did not test this hypothesis yet, we conjecture that this bias correction method could be used to correct the bias in correlation networks as well.

The $\chi$ network assesses extremal dependence integrated over time, which is useful for assessing risk. However, as such it cannot be used to explore the ``dynamics'' (i.e., year-to-year) structure of extreme events in order to gain insights about the potential mechanisms that contribute to extreme behavior, such as the 2017 hurricane season. In Section \ref{Sec3} we develop another network type, the annual extremal network, 
that aims to explore the potential drivers of extreme behavior.

In Section \ref{Sec4}, we create a $\chi$ network to hurricane-season maximum daily precipitation data collected in the US Gulf Coast to explore the spatial pattern of the extremal network connections. We also apply the annual extremal network to this precipitation data and examine whether sea surface temperature affects characteristics of the annual networks.

It should be noted that the purpose of the new network types proposed here is not to {\it replace} existing network types, such as correlation networks or event synchronization networks, but to supplement them. Namely, for any given application one might construct both a correlation network, which is more sensitive to identifying dependencies near the mean of the distribution, and a $\chi$ network, which is more sensitive to identifying dependencies between extremes. The two different networks are thus expected to provide two complementary viewpoints of the spatial dependencies. However, comparing these types of networks is not trivial, so is subject of future research, and here we simply focus on presenting the new type of network.

\section{$\chi$ and Annual Extremal Networks} \label{Sec2}

In this section we first describe how to construct a $\chi$ network, but also in tandem describe an illustrative simulation study.
Our reason for interweaving these two discussions is that the simulation study exposes a bias issue with the $\chi$ network that arises due to the act of thresholding at a level of $\chi$ and creating edges for pairs of nodes whose estimated $\hat \chi$ exceeds this threshold. We will begin by introducing the needed background in order to describe the procedure for estimating $\chi$ for each pair of nodes/locations in Section~\ref{Sec2.2}. We then present a simulation study to illustrate the issue of network bias (Sec.~\ref{Sec2.1}), which we then correct by conducting a spatial regularization on $\chi$ (Sec.~\ref{Sec2.4}). We conclude the section by describing the annual extremal networks (Sec.~\ref{Sec3}).  

\subsection{An Empirical Estimator of $\chi$} \label{Sec2.2}
We first discuss how we estimate $\chi$ from observations of a random sample of a bivariate random vector. To develop an estimator for $\chi$ one would need to first estimate the marginal distributions. Second, since $\chi$ is the limiting (conditional) probability, one would need to choose a threshold $u \in (0, 1)$ to estimate $\chi$.
There is a bias-variance trade-off to be made in the choice of $u$. 
Using a $u$ that is not sufficiently high results in increased bias, while using a $u$ that is too high results in increased variance due to paucity of data. 
In practice, often $\hat \chi (u)$ is estimated for many values of $u$ and plotted (see, for example, the {\tt chiplot} function of the {\tt evd} package \citep{stephenson2002} for {\tt R} \citep{R}).


Our particular application, with an additional assumption of max-stability, allows us to avoid the selection of $u$.
Our motivation here is to better understand relationships between hurricane events in a given year, which leads us to investigate annual extremes.
To make this precise, let\ $\bm{z}_{t} = (z_{1,t}, z_{2,t})^{\text{T}} = \left(\max_{i=(t-1)*b +1}^{b*t}(y_{1,i}), \max_{i=(t-1)*b +1}^{b*t}(y_{2,i})\right)^{\text{T}}, t = 1, \cdots, m = n/b$, where $\bm{y_{i}} = (y_{1,i}, y_{2,i})^{\text{T}},  i = 1, \cdots, n$ be the original observations with sample size $n$, $b$ be the block size, and $m = n/b$ be the sample size for the block maximum. For example, $\{\bm{y}_{i}\}_{i=1}^{n}$ could denote the daily rainfall amounts at two locations, and $i$ 
the time index (e.g., starting from January 1st for some year). If we use $b = 365 $ (or $366$ for leap years), then $\bm{z}_{t}$ would be the annual maximum precipitation amounts at these two locations in year $t$, where their dependence is of interest for risk assessment.


Below, we use estimators of extremal dependence that have been proposed for the max-stable setting to obtain our estimator for $\chi$.
Max-stability is an important notion that underlies much of extreme value analysis and here we give a brief introduction, The reader is referred to \citep{coles2001,beirlant2004,de2006,segers2012} for more details. A multivariate distribution $\bm{Z} \in \mathbb{R}^{d}$ is max--stable if the distribution is close under (component-wise) maximization. That is, each marginal distribution, by \textit{Extremal Types Theorem} \citep{fisher1928,gnedenko1943}, is belongs to the family of \textit{generalized extreme value (GEV)} \citep{jenkinson1955,coles2001}. Furthermore, the dependence structure, which can be described by its copula $C_{*}$ \citep{joe1997,nelsen2007} is satisfied the property that $C_{*}(u_{1}^{1/N}, \cdots, u_{d}^{1/N})^{N} = C_{*}(u_{1}, \cdots, u_{d})$ for all $N>0$. The copula $C_{*}$ is called \textit{max-stable copula} which form the basis for much of multivariate and spatial extreme value analysis \citep{davison2012,ribatet2013}. The implication of max-stability is that any pairwise $\chi$ for a max-stable random vector is a constant with respect to $u$ and hence there is no need to choose a threshold. Moreover, there exists deterministic relationships between $\chi$ and the other aforementioned summary statistics for extremal dependence, namely, $\theta$ and $\nu$ under the max-stable condition. We will make use of these relationships to develop our estimator for $\chi$ in the next paragraph.


We first estimate the marginal distributions of $Z_{1}$ and $Z_{2}$, denoted as $G_{1}, G_{2}$, non-parametrically by their empirical distribution functions (EDFs) $\hat{G}_{1}, \hat{G}_{2}$. We then estimate the F-madogram, the first order absolute difference in CDF 
\begin{equation}
\nu = \frac{1}{2}\mathbb{E}(|G_{1}(Z_{1})-G_{2}(Z_{2})|),
\end{equation}   
using the following empirical estimator
\begin{equation}
\hat{\nu} = \frac{1}{2}\frac{1}{m}\sum_{t=1}^{m}(|\hat{G}_{1}(z_{1,t})-\hat{G}_{2}(z_{2,t})|).
\end{equation}
\citep{cooley2006} has shown that there is a one to one correspondence between $\nu$ and $\theta$
\begin{equation}
\theta = \frac{1+2\nu}{1-2\nu}.
\end{equation}
Furthermore, by the max-stability assumption, we have $\chi = 2 -\theta$. Therefore we use the plug-in estimate $\hat{\theta} = (1+2\hat{\nu})/(1-2\hat{\nu})$ to obtain the resulting estimator of $\chi$:
\begin{equation}
\hat{\chi} = 2 - \frac{1 + 2 \hat{\nu}}{1 - 2 \hat{\nu}}.
\end{equation}

To create a $\chi$ network, we first apply this estimator to all the pairwise components $\{(i,j)\}$ of a $d$ dimensional random vector $\bm{Z}$. We then connect those pairs with $\{\hat{\chi}_{ij}\}_{i=1,\cdots, d}^{j=1,\cdots, d}, i \neq j$ greater than a chosen threshold $\chi_{min}$ to construct the corresponding $\chi$ network.

\subsection{Simulation Study} \label{Sec2.1}

The purpose of this simulation study is to examine the performance of the empirical estimator introduced in the previous section for estimating the $\chi$ network, where the network consists of pairs with $\chi$ greater than some pre-specified threshold. To do that we will need to generate data from a random vector for which we know the true network for a given threshold. The estimated network can then be compared with the true network for evaluation. This simulated network thus serves as a benchmark for the proposed new network estimation procedure. 

To make this simulation relevant in the context of environmental applications, we first simulate 100 spatial locations uniformly in a $[0,1] \times [0,1]$ domain and these are the nodes of the network. We then generate realizations from a stationary and isotropic Brown-Resnick process \citep{brown1977,kabluchko2009} at these locations. A Brown–Resnick process is a flexible class of stationary max-stable processes where every finite collection of those random variables has a multivariate max-stable distribution. The dependence structure of a Brown–Resnick process is induced by a Gaussian process via the spectral representation of max-stable process \citep{de1984}. The interested reader is referred to \citep{kabluchko2009,huser2013,thibaud2016} for more details. The implied $\chi$ value of a Brown–Resnick process for each pair $(i,j)$ only depends on their spatial distance, denoted by  $h_{ij}$, and the (Gaussian) process parameters $\rho$, the range, and $\kappa$, the smoothness: $\chi(h) = 2 - 2\Phi(\sqrt{\frac{h^{\kappa}}{2\rho }})$, where $\Phi$ is the CDF of the standard normal distribution. 
Setting $\rho = 0.05$, $\kappa = 1$, and $\chi_{min} = 0.3$, two nodes will be connected if their distance is less than 0.107, a relatively small distance compare with the spatial domain in our simulation study.

\begin{figure}[H] 
\centering
\includegraphics[width=6in]{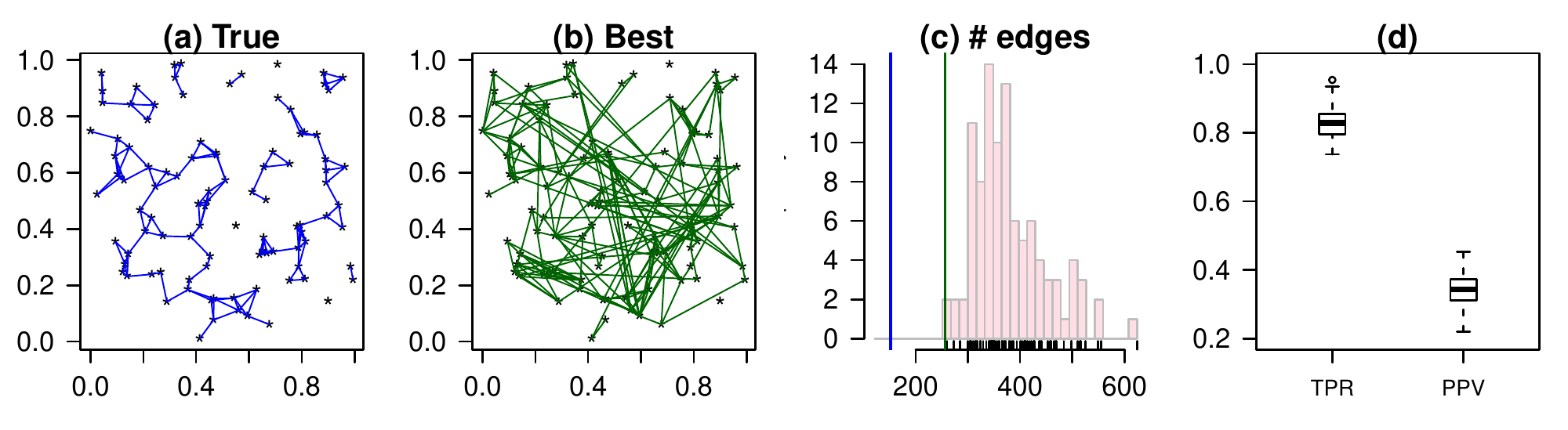}
\caption{\textbf{(a)}: The true $\chi$ network. \textbf{(b)}: The best network estimate (i.e., the number of node connections is closest to the true network). \textbf{(c)}: Histogram of estimated number of node connections, the blue and green vertical lines indicate the true and the ``best'' estimated value of the connections, respectively. (d) The TPR and PPV of the $\chi$ network connections.}
\label{fig:net1}
\end{figure}

We simulate 100 realizations (the Monte Carlo sample size), each with sample size $m = 50$ and dimension $d = 100$ (i.e., a Brown-Resnick process observed at the aforementioned 100 spatial locations). For each realization in the simulation study, we estimate the $\chi$ network with $\chi_{min} = 0.3$
using the empirical estimator described in Sec.~\ref{Sec2.2}. Comparing the estimated networks and the true network, it is apparent that the number of edges is overestimated 
(see Fig.~\ref{fig:net1} (a), (b) and (c)). We further evaluate the estimation performance in terms of the \textit{true positive rate} (TPR) and \textit{positive predictive value} (PPV), defined as

\begin{equation}
    \text{TPR} = \frac{\#\{(i,j): \hat{\chi}_{ij} >0.3 \text{ and } \chi_{ij} > 0.3\} }{\#\{(i,j): \chi_{ij}>0.3\}};
\end{equation}

\begin{equation}
    \text{PPV} = \frac{\#\{(i,j): \hat{\chi}_{ij} >0.3 \text{ and } \chi_{ij} > 0.3\} }{\#\{(i,j): \hat{\chi}_{ij} > 0.3\}}.
\end{equation}
While TPR is reasonably high ($\sim$ 80 \%), the estimation procedure identifies many more false positives than true positives as the PPV is rather low ($\sim 35 \%$, see Fig.~\ref{fig:net1} (d)). 

\begin{figure}[H] 
\centering
\includegraphics[width=3.75in]{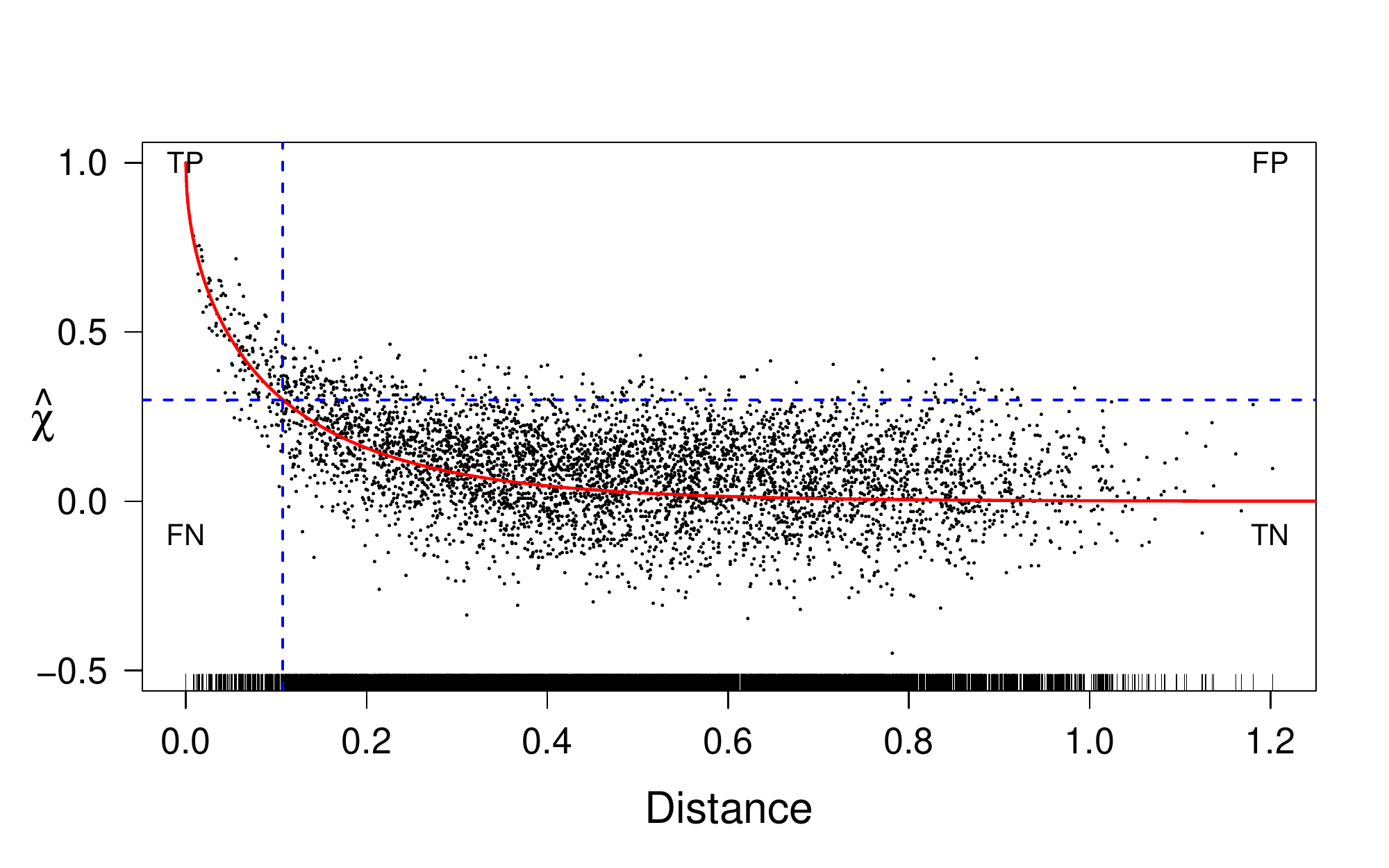}
\caption{$\{\hat{\chi}_{ij}\}$ as a function of spatial distance. The red curve is the true $\chi$ function. The black points are individual estimates $\{\hat{\chi}_{ij}\}$ from one realization. The horizontal blue dashed-line is the cutoff ($\chi_{min} = 0.3$) of the $\chi$ network and the vertical blue dashed-line is the spatial distance cutoff of the true $\chi$ network. The texts \textbf{TP, FP, FN, TN} denote true positive, false positive, false negative, and true negative, respectively.}
\label{fig:net2}
\end{figure}

Fig.~\ref{fig:net2} provides insight into how this biased estimate of the number of connections arises.
Plotted is the true $\chi$ as a function of spatial distance along with one simulation's $\chi$ estimates.  Although the estimates themselves appear unbiased as they ``center'' around the true values, it is the act of thresholding which introduces the bias.
Clearly there are more false positives than false negatives due to the shape of the true $\chi$ function and the increasing variance of the estimates as $\chi$ decreases. Note that this problem is not unique to the $\chi$ networks proposed here. In fact, we confirmed in another simulation that the same effect can also occur for the well-established correlation networks, although, to the best of our knowledge, this issue has never been mentioned in the literature on climate networks.


\subsection{Network Bias Correction} \label{Sec2.4}

Here we develop a method to correct the bias of the network estimation as demonstrated in the previous subsection. 
We exploit the spatial structure revealed from $\{\hat{\chi}_{ij}\}$ (see Fig.~\ref{fig:net2}) to spatially regularize the network estimate. In what follows we will describe a new estimator for $\{\chi_{ij}\}$, denoted by $\{\tilde{\chi}_{ij}\}$, which is a weighted average of empirical estimator $(\{\hat{\chi}_{ij}\})$ and an estimator that explicitly incorporates spatial information $(\hat{\chi}(h_{ij}))$. The weight for a given pair $(i,j)$, denoted as $\lambda_{ij}$, will depend on the estimation precision of $\hat{\chi}_{ij}$ and $\hat{\chi}(h_{ij})$. The resulting estimator, the weighted average of $\hat{\chi}_{ij}$ and $\hat{\chi}(h_{ij})$, has an \textit{empirical Bayes} interpretation (see \cite{loader1992} in the context of spatial covariance function estimation), can then be used to construct the $\chi$ network. Specifically, the bias-corrected estimator has the following form:
\begin{equation} \label{eBayesEst}
    \tilde{\chi}_{ij} = \lambda_{ij}\hat{\chi}_{ij} + (1-\lambda_{ij})\hat{\chi}(h_{ij}), \quad i, j = 1, \cdots, d.
\end{equation}

In the following we first describe how to estimate $\{\chi(h_{ij})\}$ and then discuss how to come up with the weights $\{\lambda_{ij}\}$.

\subsubsection{The estimator for $\chi(h)$}
We smooth the $\{\hat{\chi}_{ij}\}$, the empirical estimates for all the pairs, by spatial distance to obtain $\hat{\chi}(h)$. Specifically, we divide $\{\hat{\chi}_{ij}\}$ into $K$ distance bins and we compute the mean value of $\{\hat{\chi}_{ij}\}$ within each bin, denoted by $\hat{\chi}^{k}$. We then fit a cubic smoothing spline regression \citep{wahba1990,gu2013} to $\{h^{k}, \hat{\chi}^{k}\}_{k=1}^{K}$, where $h^{k}$ denotes the average distance within the $k_\text{th}$ bin, using the \texttt{sreg} function in the \texttt{fields} \texttt{R} package \citep{fieldsR} to get the $\hat{\chi}(h)$ as a function of spatial distance only. Note that this approach effectively introduced a spatially stationary and isotropic structure on $\chi$, which is less tenable especially when $h$ is ``large''. Therefore, we assume for any $h>0$, $\hat{\chi}(h)$ is normally distributed with variance $\tau^{2}(h)$ increases with spatial distance in order to weaken this somewhat strong and rigid assumption in order to learn potential nonstationary pattern on $\chi$.    

\subsubsection{The weights $\{\lambda_{ij}\}$}
The weight $\lambda_{ij}$ for a given pair $(i,j)$ is determined by the estimation variability of $\hat{\chi}_{ij}$ ($\sigma^{2}_{ij}$), and the amount of departure from $\hat{\chi}(h)$, quantified by a variance function:  $\tau^{2}(h)$. We perform a ``spatial'' block bootstrap (see e.g., \cite{wang2016}, Appendix C, or \cite{huang2016}, Appendix D) by re-sampling the vectors $\{\bm{z}_{t}\}_{t=1}^{m}$ to obtain the bootstrap variances $\{\sigma^{2*}_{ij}\}$ as the estimates for the corresponding $\{\sigma^{2}_{ij}\}$. We then use $\tau^{2}_{ij}/(\tau^{2}_{ij}+ \sigma^{2*}_{ij})$ as the weight $\lambda_{ij}$. 

The estimator (\ref{eBayesEst}) can be formulated as the posterior mean from the hierarchical model
\begin{align}
     & \hat{\chi}_{ij} \sim \text{N}(\chi_{ij}, \sigma^{2}_{ij});\\
     & \chi_{ij} = \hat{\chi}(h_{ij}) + \omega_{ij};\\
     & \omega_{ij} \sim \text{N}(0, \tau^{2}(h_{ij})).
\end{align}
$\tilde{\chi}_{ij}$ can be interpreted as an (empirical) Bayes \citep{robbins1955,efron1972,casella1985} estimator, where the empirical aspect arises from the fact that we estimate $\chi(h)$ from $\{\hat{\chi}_{ij}\}$.

\subsubsection{An illustration} 

\begin{figure}[H] 
\centering
\includegraphics[width=6.5in]{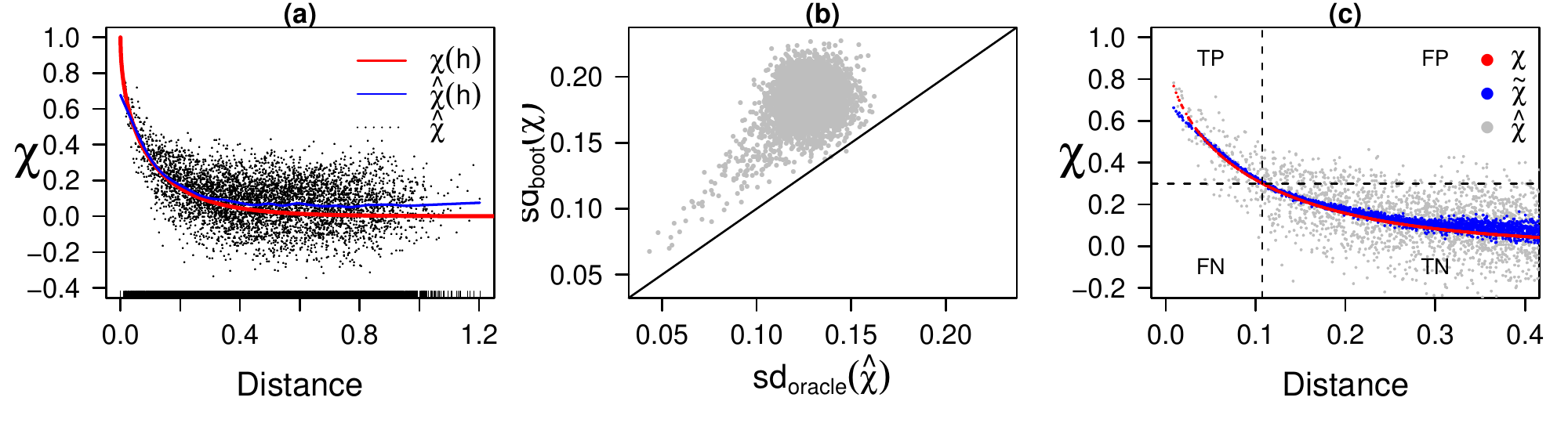}
\caption{Illustration of bias correction. \textbf{(a)}: The estimated $\chi(h)$ (blue curve), true $\chi(h)$ (red curve), and the empirical estimate $\hat{\chi}$ (black points). \textbf{(b)}: The scatter plot of $(\sigma_{oracle, ij}, \sigma_{boot, ij})$. All the points are above the 1:1 line means that the bootstrap standard errors are slightly higher than the ``true'' standard errors. \textbf{(c)}: As in Fig.~\ref{fig:net2} but we add the bias-corrected estimates $\{\tilde{\chi}_{ij}\}$ (blue points).}
\label{fig:bc_sim}
\end{figure}

Following the bias correction procedure described in previous subsections, we first bin the $\{\hat{\chi}_{ij}\}$ into $K = 100$ bins to obtain $\{\hat{\chi}^{k}\}_{k=1}^{100}$. We then fit a cubic spline regression to $\{h^{k}, \hat{\chi}^{k}\}_{k=1}^{100}$ to get the estimate of $\chi(h)$ (see Fig.~\ref{fig:bc_sim}, (a)), we then \textit{assume} $\tau^{2}(h) = 0.095/\left(1 + \exp\left(-6 * \left(h - 0.72\right)\right)\right)$, which is an increasing (logistic) function with $h$. As described in previous subsection, the bootstrap standard deviations are used as the estimates of $\{\sigma_{ij}\}$ and we compare those estimates with the ``oracle'' standard errors (i.e., standard deviations (SDs) among those 100 Monte Carlo simulations from the true model). As would be expected the bootstrap SDs are (slightly) higher than their corresponding oracle SDs but they show similar behavior, at least qualitatively. We can then plug in the $\sigma^{2*}_{ij}$, $\hat{\chi}(h_{ij})$, and $\tau^{2}(h_{ij})$ to obtain the resulting estimate $\tilde{\chi}_{ij}$ for each pair to construct the bias-corrected network. This estimator performs well as $\{\tilde{\chi}_{ij}\}$ have been ``shrunk'' toward $\hat{\chi}(h_{ij})$, which is much closer to the truth than $\{\hat{\chi}_{ij}\}$ (Fig.~\ref{fig:bc_sim}, (c)).            

The comparison of TPR/PPV for the empirical and the bias-corrected $\chi$ network estimators shows that our method not only identifies most of the true connections, i.e., TPR $\approx$ 1, but also avoids making many false discoveries, i.e., PPV for the bias-corrected estimator is substantially higher than that of the empirical estimator (see Tab.~\ref{table1}).

\begin{table}[H]
\caption{Summary of TPR/PPV for the empirical $\hat{\chi}$ and the bias-corrected $\tilde{\chi}$ network (i.e., $\chi$ > 0.3) estimators.}\label{table1}
\begin{center}
\begin{tabular}{ c c c c c c } 
 \hline
 TPR percentile &  5th  & 25th & 50th &  75th & 95th \\ 
 \hline
 \hline
$\hat{\chi}$ & 0.76 & 0.80 & 0.83& 0.86& 0.91\\
$\tilde{\chi}$ & 0.88 & 0.97 & 1 & 1 & 1\\
 \hline
 PPV percentile &  5th  & 25th & 50th &  75th & 95th \\ 
 \hline
 \hline
$\hat{\chi}$ & 0.26 & 0.31 & 0.34& 0.37& 0.41\\
$\tilde{\chi}$ & 0.58 & 0.76 & 0.88 & 0.99 & 1\\
 \hline
\end{tabular}
\end{center}
\end{table}

This bias correction method, with some modifications, is expected to apply also to correlation networks to compensate for their network bias, but that topic is beyond the scope of this work.


\subsection{Annual Extremal Network} \label{Sec3}
The estimation procedure of the $\chi$ network and its bias-corrected version assumes that the dependence structure of the block maximum, $\{\bm{Z}_{t}\}$, does not change with ``time'' (i.e., $\bm{z}_{t}$, $t = 1, \cdots, m$ is a random sample from an underlying time invariant random vector $\bm{Z}$), so that one can use these ``replicates'' to infer the spatial extremal dependence. Moreover, we assume that $\bm{Z}$ is max-stable to avoid the need to choose 
a high threshold $u$ to estimate $\chi$. Here we introduce another type of network, the annual extremal network, 
to examine the year-to-year variation of the ``observed'' extremal networks and we seek to explain this variability. The main idea of the annual extremal network is to connect a pair of nodes in a given ``year'' $t$ if extreme events occur at both locations. For example, an extreme event can be defined as the value exceeding 
a high percentile, say the 95th percentile. In this case, two locations are connected if both locations exceed their 20-year return level in a given ``year''. Specifically, we treat the EDFs of $\{\bm{z}_{t}\}$, denoted by $\{\bm{u}_{t}\}$, where $\bm{u}_{t} = \{u_{1,t}, u_{2,t}, \cdots, u_{d,t}\}^{T}$, as the ``data'' and for each $t = 1, \cdots, m$, we connect pairs $(i,j)$ if both $u_{i,t}$ and $u_{j,t}$ are greater than $0.95$. 

The definition of the annual extremal network defined here has similaraties to the definition of event synchronization networks \citep{malik2010spatial,malik2012analysis,boers2014south}, but the latter operate on a completely different time scale.
Event synchronization networks in climate science typically consider daily data, and define extreme events at each location to be those exceeding a certain threshold, e.g.\ the 95th percentile of the overall distribution. Two locations are connected if an extreme event at one location is frequently followed by an extreme event at another location within a matter of {\it days}, which can be used to track the {\it propagation} of individual phenomena from one location to another, e.g., the propagation of storm systems.  In contrast, the annual extremal network, due to its annual scale, investigates whether {\it the overall conditions in a given year} lead to several events in the same year, although those events may not have any {\it direct} connection to each other. Thus, due to the different time scale (days vs.\ years) these two network types focus on very different phenomena.

Summary statistics (e.g., the number of the connections, the average distance of connections at each location) of this annual network are time series (on an annual scale) and can be related to some physically meaningful covariates to explore its year-to-year variation. 
In Section \ref{Sec4.3} we use the number of ``long distance'' connected pairs to relate this statistic to sea surface temperature. Note that the annual extremal network described here is constructed directly from the ``data'', namely the EDFs for all the locations, whereas the $\chi$ network is constructed from the $\chi$ statistic that was estimated from the EDFs. Therefore, by ignoring the estimation uncertainty of EDFs themselves, we do not need to develop a bias correction procedure as we did for the $\chi$ network.        

\section{Gulf Coast Extreme Precipitation Application}\label{Sec4}

\subsection{Data Set}
We use a subset of the Global Historical Climatology Network (GHCN) data set \citep{GHCN}, where the spatial domain includes Texas, Louisiana, Mississippi, Alabama, Florida, and Georgia and the time period is from Jan.\ 1949 to Oct.\ 2017. There are 339 GHCN stations being included in this study where the selecting criterion is that those stations should have at least 90 percentage of non-missing daily precipitation values. We further restrict our attention to the annual maximum of the daily precipitation values from June to October to investigate the extremal dependence structure for the hurricane season.

\subsection{$\chi$ Network for Gulf Coast Extreme Precipitation} \label{Sec4.2}

\begin{figure}[H] 
\centering
\includegraphics[width=4.6in]{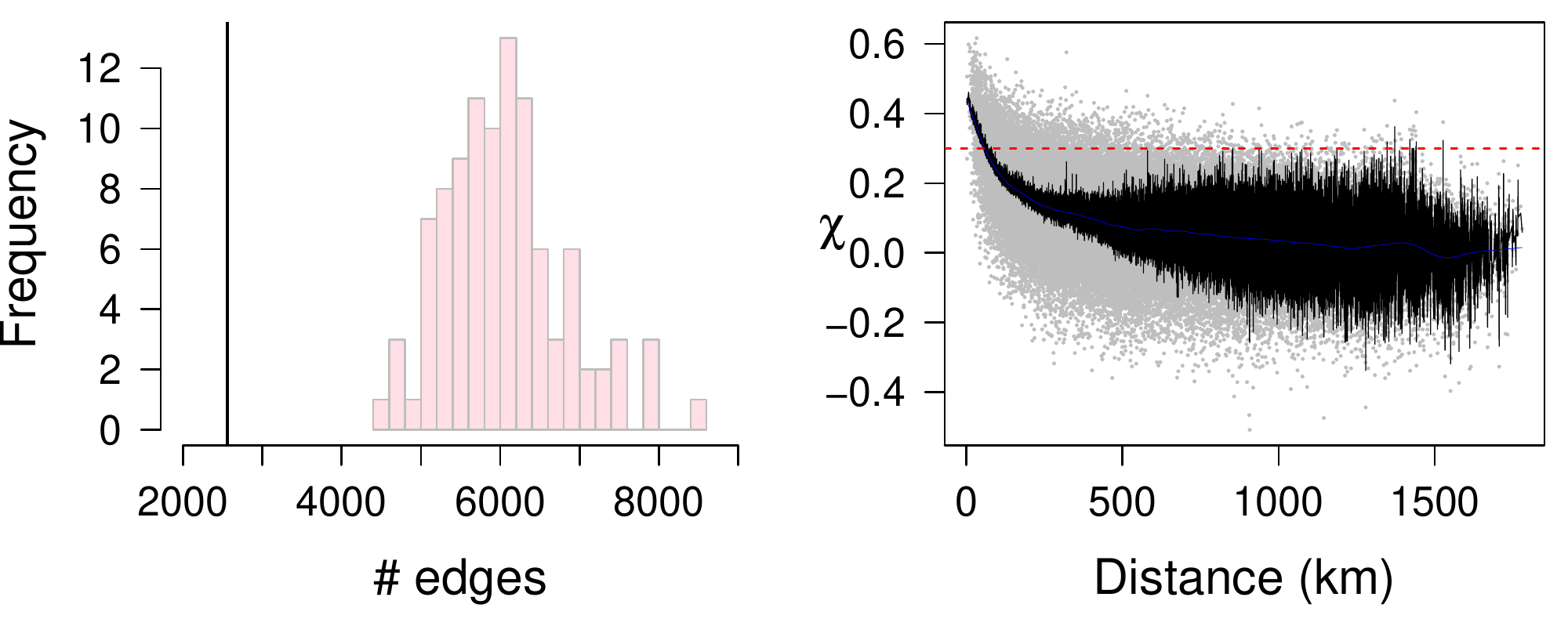}
\caption{\textbf{Left}: Histogram of estimated number of node connections of the block bootstrap sample, the black vertical lines indicate the estimated value of the connections of the original sample. \textbf{Right}: $\{\hat{\chi}_{ij}\}$ (gray points), $\chi(h)$ (blue curve), and $\{\tilde{\chi}_{ij}\}$ (black curve) of the Gulf Coast hurricane season maximum precipitation.}
\label{fig:bc_gulf}
\end{figure}

We first compute the $\chi$ network based on the empirical estimator as described in Sec.~\ref{Sec2.2} and we assess the network estimation uncertainty using a spatial block bootstrap. The results obtained from the block bootstrap (Fig.~\ref{fig:bc_gulf}, Left panel) clearly demonstrate the issue of overestimation. We then conduct a bias correction using the method described in Sec.~\ref{Sec2.4}. The right panel in Fig.~\ref{fig:bc_gulf} displays the $\{\hat{\chi}_{ij}\}$, $\chi(h)$, and $\{\tilde{\chi}_{ij}\}$ as a function of the spatial distance.

\begin{figure}[H] 
\centering
\includegraphics[width=4.4in]{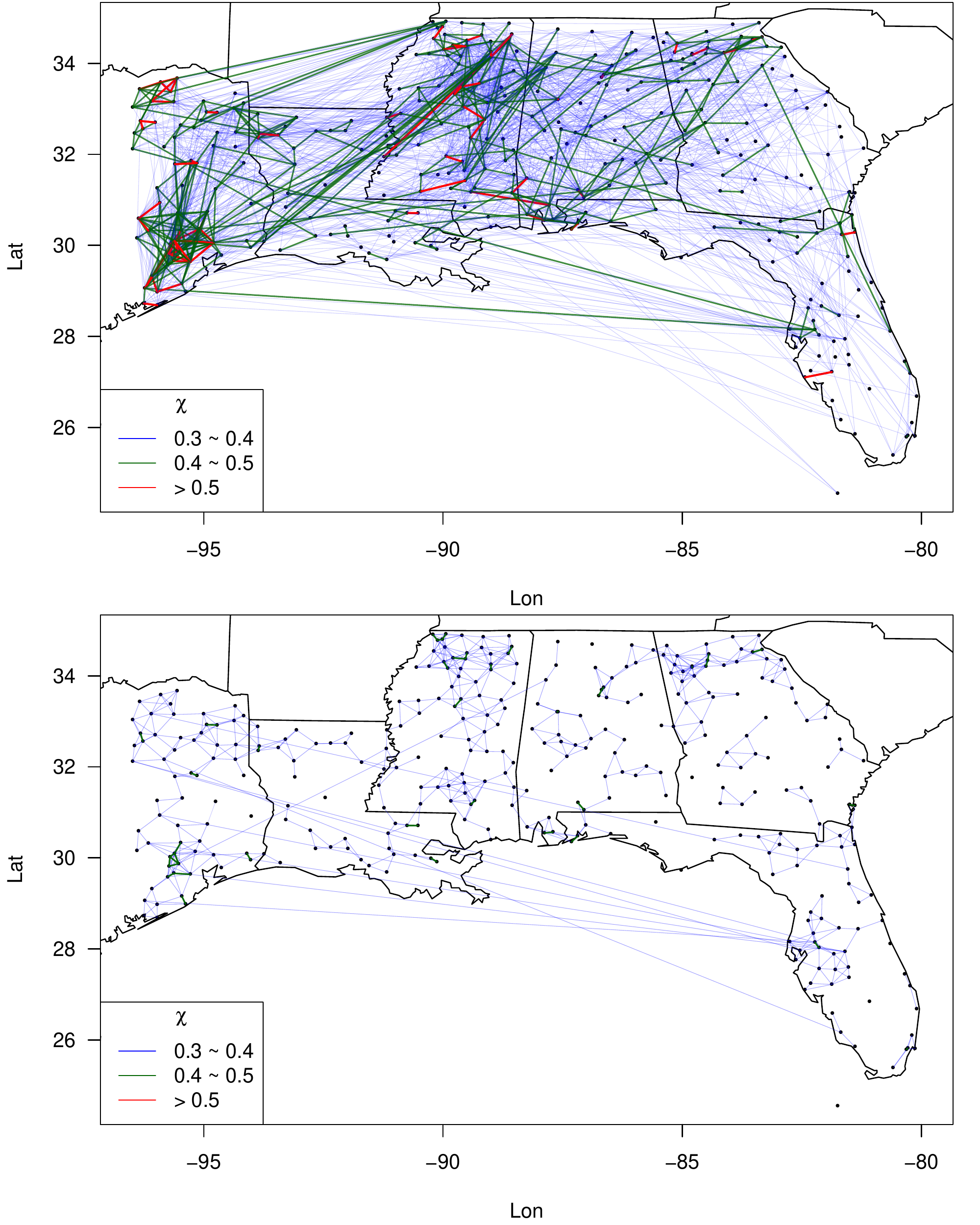}
\caption{$\chi$ network of the hurricane-season maximum daily precipitation across US Gulf Coast. \textbf{Top}: The initial empirical estimate of the $\chi$ network. \textbf{Bottom}: The bias-corrected $\chi$ network estimate. (The connections in this figure are best viewed on-screen.)}
\label{fig:chi_network}
\end{figure}

Fig.~\ref{fig:chi_network} shows that many long distance links in the $\chi$ network  based on the empirical estimator disappear and may simply be due to estimation variability for using the empirical estimator. However, the bias-corrected $\chi$ network does identify some long distance edges, for example between the Houston and Tampa regions. It is important to note that although $\chi$ takes on values between 0 and 1, its values should not be interpreted in a similar manner to correlation values.  A $\chi$ of 0.3 between two distant locations would imply strong tail dependence; that is, when one location experiences a very extreme event, the other would have a 30 percent chance of also experiencing an extreme event.  Individual links are hard to interpret, and we do not believe there is some phenomenon that specifically links Houston to Tampa.  Nevertheless, these long-distance network links at a minimum suggest that calculating annual risk between two distant regions like Houston and Tampa should not be done with an independence assumption as this would  underestimate joint risk.

\subsection{Annual Extremal Precipitation Network for Gulf Coast} \label{Sec4.3}
\begin{figure}[H] 
\centering
\includegraphics[width=6in]{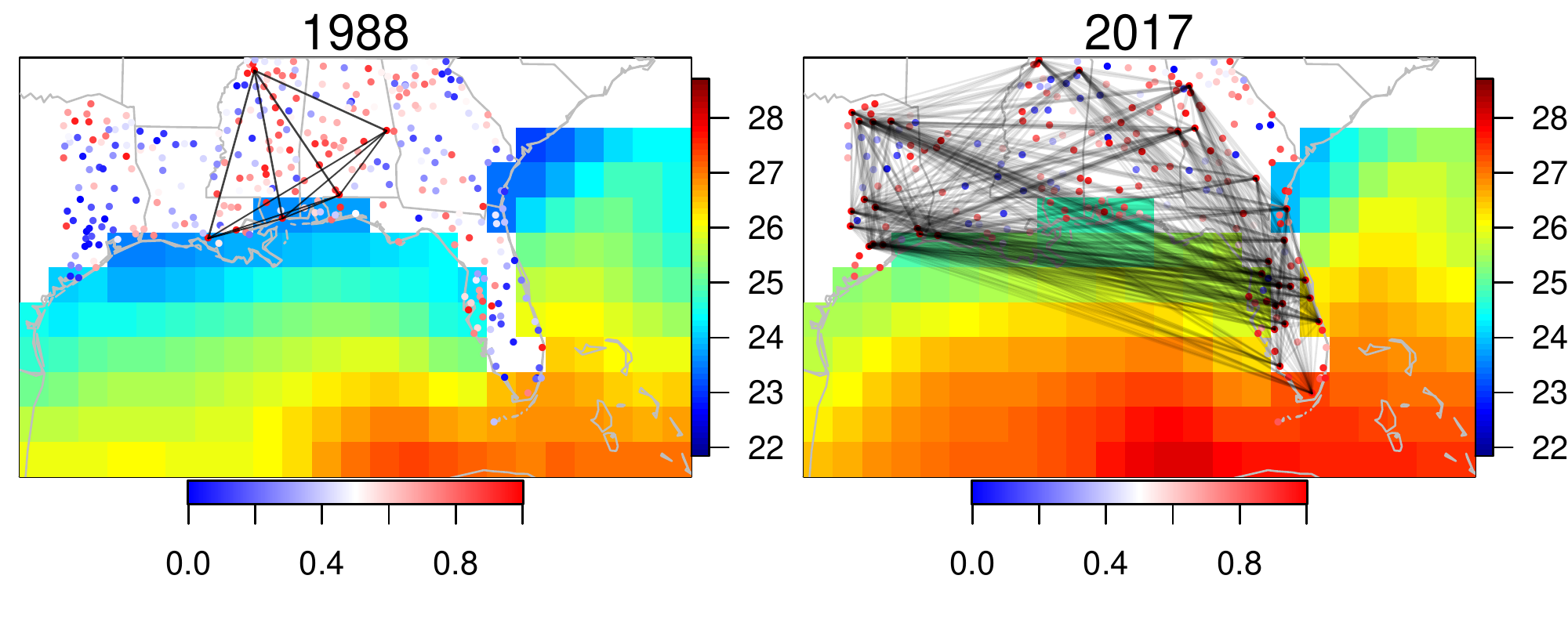}
\caption{Annual extremal precipitation networks for the year 1988 (\textbf{Left}) and 2017 (\textbf{Right}). The EDFs are plotted (circles) in the spatial domain using a blue-white-red color scheme (see the horizontal color bars). These two annual extremal networks illustrate the year to year variation. Note that the year 1988 is the year with lowest Gulf Coast SST and 2017 is the year with the highest SST. SST are added as images using a rainbow color scheme (see the vertical col bars).}
\label{fig:ann_net}
\end{figure}

In this subsection we construct annual extremal networks in the study region to examine the year-to-year structure. Specifically, for each hurricane season we identify the connections where their EDFs of the hurricane season maximum exceed 0.95 (i.e., both places experience at least a 20-year event in that year, see Fig.~\ref{fig:ann_net} for examples). To further sutdy the long range extremal dependence, we compute the number of pairs where their spatial distance are greater than 1,000km. We then study how the inter-annual variability of the numbers of these ``long distance'' extremal pairs might be explained by some meteorological covariate, for example, annual mean sea surface temperature (SST) averaged in a relevant spatial region. Here we use the Hadley Centre Global monthly average Sea Surface Temperature version 1.1 (HadISST 1.1, \cite{HadISST2003}) and we compute the spatial average of the annual mean SST over a ``rectangular'' region with longitude from $95^{\circ}$ west to $83^{\circ}$ west and latitude from $23^{\circ}$ north to $29^{\circ}$ north where the annual window is defined as from July of the previous year to June (i.e., the 2017 mean SST is computed by averaging the monthly SST from July 2016 to June 2017).

\begin{figure}[H] 
\centering
\includegraphics[width=6in]{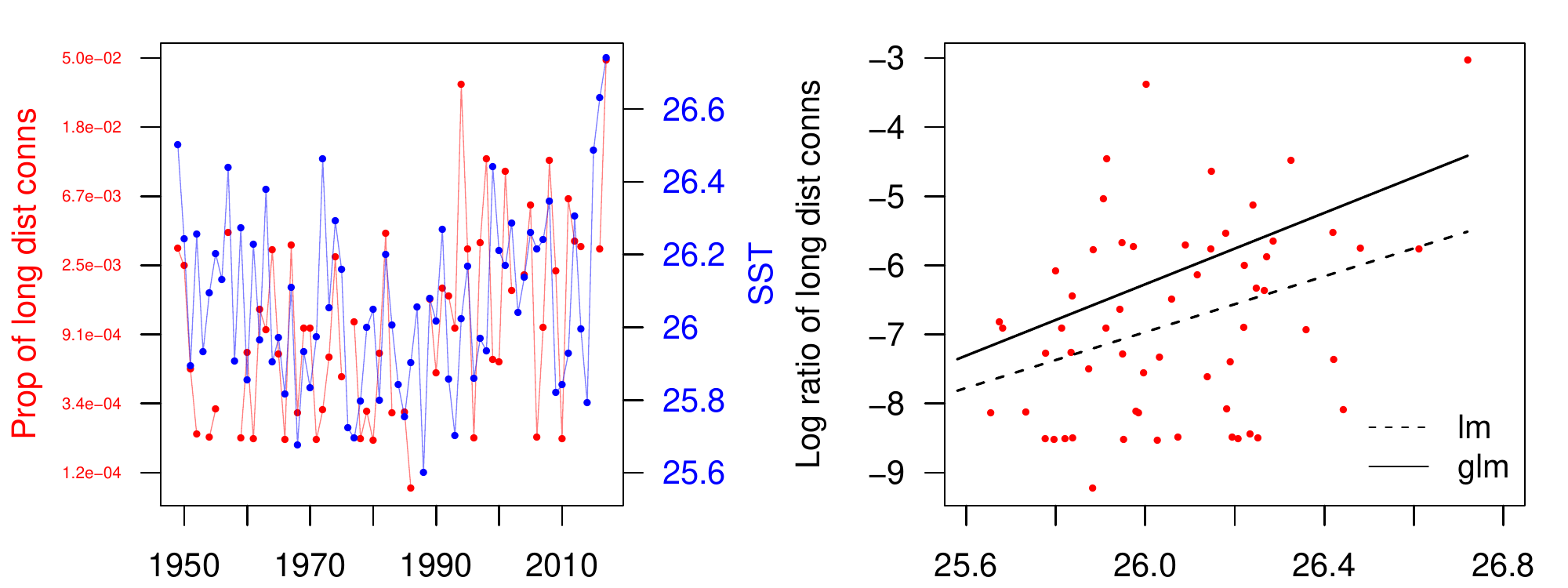}
\caption{\textbf{Left}: The time series of the proportion of the number of long distance extremal connections (red) and the time series of Gulf Coast annual mean SST (blue). \textbf{Right}: Scatterplot of the log of the long distance extremal connections versus Gulf Coast SST. The dashed line represents the linear regression (lm) fit and the solid line represents the Poisson regression (glm) fit.}
\label{fig:yearly_net}
\end{figure}

We explore the potential cause of the annual variation in the number of extremal connections as shown in Fig.~\ref{fig:ann_net}. In Fig.~\ref{fig:yearly_net} (left panel), we plot two time series together, the red one for the proportion of the long distance (i.e., > 1,000km) extremal connections on a log scale (log ratio hereafter), where this log ratio is computed as the logarithm of the number of the connections divided by the total number of the long distance pairs in the study region in a given year. The blue line is the SST time series. There exists ``co-movement'' between these two time series motivated us to perform regression analyses where we use the log ratio as the response and SST as the predictor (see Fig.~\ref{fig:yearly_net}, Right panel).
Specifically, we perform two regression analyses to study this relationship. First, a linear regression (\texttt{lm}): $$\text{log ratio} = \alpha_{0} + \alpha_{1}\text{SST} + \varepsilon.$$ Second, a generalized linear model \citep{mccullagh1989} assuming the number of the long distance extremal connections  follows a Poisson distribution with its log intensity, denoted by $\log(\mu)$, depends linearly on SST (\texttt{glm}): $$\mathbb{E}[\log(\mu)|\text{SST}] =  \beta_{0} + \beta_{1}\text{SST}.$$ Both the \texttt{lm} (although the \textit{homoscedasticity} assumption is likely not appropriate here) and \texttt{glm} model fits suggest that the SST can explain some proportion of the year-to-year variation observed in the number of long distance extremal pairs, and the estimated slopes ($\hat{\alpha}_{1}$ and $\hat{\beta}_{1}$) are positive with p-values $0.008$ and $<2 \time 10^{-16}$. The results here suggest that the annual mean Gulf Coast SST may partly contribute to the abnormally long distance extremal dependence observed in the 2017 hurricane season.

\subsection{Scientific Implications} \label{Sec4.4}

More intense hurricanes (i.e.,\ named storms) under climate change lead to great US economic losses \citep[e.g.,][]{emanuel2011,klotzbach2018}. Studies already found that number of extreme precipitation events associated with hurricane has increased in recent decades \citep[e.g.,][]{kunkel2010}. But it is still largely unknown how the spatial dependence of extreme precipitation may respond. Sparse summer convective precipitation events have typical radius around 10-100km, which lead to short-distance spatial dependence. However, major Atlantic hurricanes can grow to 800km measured with the outer radius of vanishing wind \citep{chavas2016}. Such large low-pressure system deliver huge amount of water when it moves from the warm ocean to land. In this case, faraway places like some regions in Texas and Florida could all encounter extreme precipitation during a hurricane season. Here our biase-corrected $\chi$ network (see Fig.~\ref{fig:chi_network}, Bottom panel) captures a range of precipitation spatial dependence, not only the common short-distance connectivity among nearby stations but also the long-distance connectivity across states associated with hurricane events. With this overall understanding for the spatial extremal dependence, we further examine the temporal variation of the annual extremal precipitation networks and we investigate the potential relationship with the US Gulf Coast SST. The analyses in Sec.~\ref{Sec4.3} reveal there exists a robust relationship between the number of long distance extremal precipitation connection in southern states of US and pre-hurricane season (previous year July to current year June) SST in Gulf of Mexico. It agrees with the understanding that increased ocean heat content leads to more tropical cyclone \citep{trenberth2018}, which in turn induces more long-distance preciptation extremes. In summary, by combining the extremal dependence estimates with the network analysis, we are able to quantitatively measure how the hurricane-induced extreme precipitation changes under climate change. The network perspective enables us to reveal the structural change among a suite of stations, which provide additional information to standard analysis based on individual locations.

\section{Discussion}\label{sec5}

This work presents two new tools, the $\chi$ and annual extremal networks, for exploring dependence structure of extreme values for space-time data sets. In contrast to the widely used correlation network in the climate literature, the $\chi$ network uses $\chi$ statistics to summarize {\it extremal} dependence, and is hence suitable for studying extremal dependence.
We demonstrate through a simulation study that a $\chi$ network constructed by using an empirical estimator could overestimate the degree of the $\chi$ network. We found that this issue can also arise if one uses the sample correlation to construct a correlation network.
We propose a bias-correction method to incorporate the spatial dependence structure to alleviate the network estimation bias for the $\chi$ network. We expect that this bias correction procedure could also be used to overcome the bias in traditional correlation networks. We then apply the $\chi$ network to the Gulf Coast hurricane season rainfall extremes to study the spatial dependence structure of the annual maximum precipitation observed over the Gulf Coast.

We also apply the annual extremal network to further explore the potential non-stationary long-range extremal dependence. Our results suggest that the sea surface temperature may contribute to the abnormally long distance extremal dependence observed in the 2017 hurricane season.

Due to the exploratory nature of the $\chi$ network, there are some limitations in terms of providing the whole picture of the extremal dependence structure of an environmental process of interest. First, the $\chi$ statistic is a measure of pairwise tail dependence and hence not able to describe the higher order extremal dependence. 
Second, $\chi$ is an appropriate extremal dependence metric in the case of asymptotic dependence, it is less clear about the usefulness of this measure if the environmental process of interest exhibit weakening spatial dependence as events become more extreme \citep{huser2017,wadsworth2017,huser2018,bopp2018}.

In this work, we make several simplifying assumptions for developing our bias-corrected method in order to obtain a close form expression for our estimator. One could try to relax these assumptions at the price that the computations needed for a sampling based approach, which can be substantial. Given that the number of pairs will be large for most environmental applications, we prefer a simple method such as the one we proposed here, that allows us to quickly explore the extremal dependence structure, which could provide some guidance on how to proceed the following confirmatory analysis.


\singlespacing
\bibliography{network.bib}

\begin{thebibliography}{68}
\providecommand{\natexlab}[1]{#1}
\providecommand{\url}[1]{\texttt{#1}}
\expandafter\ifx\csname urlstyle\endcsname\relax
  \providecommand{\doi}[1]{doi: #1}\else
  \providecommand{\doi}{doi: \begingroup \urlstyle{rm}\Url}\fi

\bibitem[Bahadori and Liu(2011)]{YanCI2012}
M.~T. Bahadori and Y.~Liu.
\newblock Granger causality analysis with hidden variables in climate science
  applications.
\newblock In \emph{Climate Informatics workshop (CI 2011)}, 2011.

\bibitem[Beirlant et~al.(2004)Beirlant, Goegebeur, Segers, and
  Teugels]{beirlant2004}
J.~Beirlant, Y.~Goegebeur, J.~Segers, and J.~Teugels.
\newblock \emph{Statistics of extremes: theory and applications}.
\newblock John Wiley \& Sons, 2004.

\bibitem[Boers et~al.(2014)Boers, Rheinwalt, Bookhagen, Barbosa, Marwan,
  Marengo, and Kurths]{boers2014south}
N.~Boers, A.~Rheinwalt, B.~Bookhagen, H.~M. Barbosa, N.~Marwan, J.~Marengo, and
  J.~Kurths.
\newblock The south american rainfall dipole: a complex network analysis of
  extreme events.
\newblock \emph{Geophysical Research Letters}, 41\penalty0 (20):\penalty0
  7397--7405, 2014.

\bibitem[Bopp et~al.(2018)Bopp, Shaby, and Huser]{bopp2018}
G.~P. Bopp, B.~A. Shaby, and R.~Huser.
\newblock A hierarchical max-infinitely divisible process for extreme areal
  precipitation over watersheds.
\newblock \emph{arXiv preprint arXiv:1805.06084}, 2018.

\bibitem[Brown and Resnick(1977)]{brown1977}
B.~M. Brown and S.~I. Resnick.
\newblock Extreme values of independent stochastic processes.
\newblock \emph{Journal of Applied Probability}, 14\penalty0 (4):\penalty0
  732--739, 1977.

\bibitem[Casella(1985)]{casella1985}
G.~Casella.
\newblock An introduction to empirical bayes data analysis.
\newblock \emph{The American Statistician}, 39\penalty0 (2):\penalty0 83--87,
  1985.

\bibitem[Chavas et~al.(2016)Chavas, Lin, Dong, and Lin]{chavas2016}
D.~R. Chavas, N.~Lin, W.~Dong, and Y.~Lin.
\newblock Observed tropical cyclone size revisited.
\newblock \emph{Journal of Climate}, 29\penalty0 (8):\penalty0 2923--2939,
  2016.

\bibitem[Coles et~al.(1999)Coles, Heffernan, and Tawn]{coles1999}
S.~Coles, J.~Heffernan, and J.~Tawn.
\newblock Dependence measures for extreme value analyses.
\newblock \emph{Extremes}, 2\penalty0 (4):\penalty0 339--365, 1999.

\bibitem[Coles et~al.(2001)Coles, Bawa, Trenner, and Dorazio]{coles2001}
S.~Coles, J.~Bawa, L.~Trenner, and P.~Dorazio.
\newblock \emph{An introduction to statistical modeling of extreme values},
  volume 208.
\newblock Springer, 2001.

\bibitem[Cooley et~al.(2006)Cooley, Naveau, and Poncet]{cooley2006}
D.~Cooley, P.~Naveau, and P.~Poncet.
\newblock Variograms for spatial max-stable random fields.
\newblock In \emph{Dependence in probability and statistics}, pages 373--390.
  Springer, 2006.

\bibitem[Davison et~al.(2012)Davison, Padoan, Ribatet, et~al.]{davison2012}
A.~C. Davison, S.~Padoan, M.~Ribatet, et~al.
\newblock Statistical modeling of spatial extremes.
\newblock \emph{Statistical Science}, 27\penalty0 (2):\penalty0 161--186, 2012.

\bibitem[de~Haan(1984)]{de1984}
L.~de~Haan.
\newblock A spectral representation for max-stable processes.
\newblock \emph{The annals of probability}, pages 1194--1204, 1984.

\bibitem[de~Haan and Ferreira(2006)]{de2006}
L.~de~Haan and A.~Ferreira.
\newblock \emph{Extreme value theory: an introduction}.
\newblock Springer, 2006.

\bibitem[Donges et~al.(2009)Donges, Zou, Marwan, and Kurths]{donges2009}
J.~F. Donges, Y.~Zou, N.~Marwan, and J.~Kurths.
\newblock Complex networks in climate dynamics.
\newblock \emph{The European Physical Journal Special Topics}, 174\penalty0
  (1):\penalty0 157--179, 2009.

\bibitem[Ebert-Uphoff and Deng(2014)]{ebert2014causal}
I.~Ebert-Uphoff and Y.~Deng.
\newblock Causal discovery from spatio-temporal data with applications to
  climate science.
\newblock In \emph{Machine Learning and Applications (ICMLA), 2014 13th
  International Conference on}, pages 606--613. IEEE, 2014.

\bibitem[Ebert-Uphoff et~al.(2018)Ebert-Uphoff, Huang, Mitra, Cooley,
  Chatterjee, Chen, and Wang]{CI2018_network_paper}
I.~Ebert-Uphoff, W.~Huang, A.~Mitra, D.~Cooley, S.~Chatterjee, C.~Chen, and
  Z.~Wang.
\newblock Studying extremal dependence in climate using complex networks.
\newblock In \emph{Proceedings of the 8th International Workshop on Climate
  Informatics (CI 2018)}, Boulder, CO, 2018.

\bibitem[Efron and Morris(1972)]{efron1972}
B.~Efron and C.~Morris.
\newblock Limiting the risk of bayes and empirical bayes estimators?part ii:
  The empirical bayes case.
\newblock \emph{Journal of the American Statistical Association}, 67\penalty0
  (337):\penalty0 130--139, 1972.

\bibitem[Emanuel(2011)]{emanuel2011}
K.~Emanuel.
\newblock Global warming effects on us hurricane damage.
\newblock \emph{Weather, Climate, and Society}, 3\penalty0 (4):\penalty0
  261--268, 2011.

\bibitem[Fisher and Tippett(1928)]{fisher1928}
R.~A. Fisher and L.~H.~C. Tippett.
\newblock Limiting forms of the frequency distribution of the largest or
  smallest member of a sample.
\newblock In \emph{Mathematical Proceedings of the Cambridge Philosophical
  Society}, volume~24, pages 180--190. Cambridge Univ Press, 1928.

\bibitem[Friedman et~al.(2008)Friedman, Hastie, and
  Tibshirani]{friedman2008sparse}
J.~Friedman, T.~Hastie, and R.~Tibshirani.
\newblock Sparse inverse covariance estimation with the graphical lasso.
\newblock \emph{Biostatistics}, 9\penalty0 (3):\penalty0 432--441, 2008.

\bibitem[Gnedenko(1943)]{gnedenko1943}
B.~Gnedenko.
\newblock Sur la distribution limite du terme maximum d'une serie aleatoire.
\newblock \emph{Annals of mathematics}, pages 423--453, 1943.

\bibitem[Gu(2013)]{gu2013}
C.~Gu.
\newblock \emph{Smoothing spline ANOVA models}.
\newblock Springer series in statistics, 297. Springer, New York, 2nd ed.
  edition, 2013.
\newblock ISBN 1299337511.

\bibitem[Halverson(2018)]{halverson2018}
J.~B. Halverson.
\newblock The costliest hurricane season in us history.
\newblock \emph{Weatherwise}, 71\penalty0 (2):\penalty0 20--27, 2018.

\bibitem[Huang et~al.(2016)Huang, Stein, McInerney, Sun, and Moyer]{huang2016}
W.~K. Huang, M.~L. Stein, D.~J. McInerney, S.~Sun, and E.~J. Moyer.
\newblock Estimating changes in temperature extremes from millennial-scale
  climate simulations using generalized extreme value (gev) distributions.
\newblock \emph{Advances in Statistical Climatology, Meteorology and
  Oceanography}, 2\penalty0 (1):\penalty0 79--103, 2016.
\newblock \doi{10.5194/ascmo-2-79-2016}.
\newblock URL \url{https://www.adv-stat-clim-meteorol-oceanogr.net/2/79/2016/}.

\bibitem[Huser and Davison(2013)]{huser2013}
R.~Huser and A.~C. Davison.
\newblock Composite likelihood estimation for the brown--resnick process.
\newblock \emph{Biometrika}, 100\penalty0 (2):\penalty0 511--518, 2013.

\bibitem[Huser and Wadsworth(2018)]{huser2018}
R.~Huser and J.~L. Wadsworth.
\newblock Modeling spatial processes with unknown extremal dependence class.
\newblock \emph{Journal of the American Statistical Association}, pages 1--11,
  2018.

\bibitem[Huser et~al.(2017)Huser, Opitz, and Thibaud]{huser2017}
R.~Huser, T.~Opitz, and E.~Thibaud.
\newblock Bridging asymptotic independence and dependence in spatial extremes
  using gaussian scale mixtures.
\newblock \emph{Spatial Statistics}, 21:\penalty0 166--186, 2017.

\bibitem[Jenkinson(1955)]{jenkinson1955}
A.~F. Jenkinson.
\newblock The frequency distribution of the annual maximum (or minimum) values
  of meteorological elements.
\newblock \emph{Quarterly Journal of the Royal Meteorological Society},
  81\penalty0 (348):\penalty0 158--171, 1955.

\bibitem[Joe(1997)]{joe1997}
H.~Joe.
\newblock \emph{Multivariate models and multivariate dependence concepts}.
\newblock Chapman and Hall/CRC, 1997.

\bibitem[Kabluchko et~al.(2009)Kabluchko, Schlather, De~Haan,
  et~al.]{kabluchko2009}
Z.~Kabluchko, M.~Schlather, L.~De~Haan, et~al.
\newblock Stationary max-stable fields associated to negative definite
  functions.
\newblock \emph{The Annals of Probability}, 37\penalty0 (5):\penalty0
  2042--2065, 2009.

\bibitem[Klotzbach et~al.(2018)Klotzbach, Bowen, Pielke~Jr, and
  Bell]{klotzbach2018}
P.~J. Klotzbach, S.~G. Bowen, R.~Pielke~Jr, and M.~Bell.
\newblock Continental united states hurricane landfall frequency and associated
  damage: Observations and future risks.
\newblock \emph{Bulletin of the American Meteorological Society}, \penalty0
  (2018), 2018.

\bibitem[Kretschmer et~al.(2016)Kretschmer, Coumou, Donges, and
  Runge]{kretschmer2016using}
M.~Kretschmer, D.~Coumou, J.~F. Donges, and J.~Runge.
\newblock Using causal effect networks to analyze different arctic drivers of
  midlatitude winter circulation.
\newblock \emph{Journal of Climate}, 29\penalty0 (11):\penalty0 4069--4081,
  2016.

\bibitem[Kunkel et~al.(2010)Kunkel, Easterling, Kristovich, Gleason, Stoecker,
  and Smith]{kunkel2010}
K.~E. Kunkel, D.~R. Easterling, D.~A. Kristovich, B.~Gleason, L.~Stoecker, and
  R.~Smith.
\newblock Recent increases in us heavy precipitation associated with tropical
  cyclones.
\newblock \emph{Geophysical Research Letters}, 37\penalty0 (24), 2010.

\bibitem[Loader and Switzer(1992)]{loader1992}
C.~Loader and P.~Switzer.
\newblock Spatial covariance estimation for monitoring data.
\newblock \emph{Statistics in the Environmental and Earth Sciences}, pages
  52--70, 1992.

\bibitem[L{\"u}tkepohl(2007)]{lutkepohl2007}
H.~L{\"u}tkepohl.
\newblock \emph{New introduction to multiple time series analysis}.
\newblock Springer Science \& Business Media, 2nd edition, 2007.

\bibitem[Malik et~al.(2010)Malik, Marwan, and Kurths]{malik2010spatial}
N.~Malik, N.~Marwan, and J.~Kurths.
\newblock Spatial structures and directionalities in monsoonal precipitation
  over south asia.
\newblock \emph{Nonlinear Processes in Geophysics}, 17\penalty0 (5):\penalty0
  371--381, 2010.

\bibitem[Malik et~al.(2012)Malik, Bookhagen, Marwan, and
  Kurths]{malik2012analysis}
N.~Malik, B.~Bookhagen, N.~Marwan, and J.~Kurths.
\newblock Analysis of spatial and temporal extreme monsoonal rainfall over
  south asia using complex networks.
\newblock \emph{Climate dynamics}, 39\penalty0 (3-4):\penalty0 971--987, 2012.

\bibitem[McCullagh and Nelder(1989)]{mccullagh1989}
P.~McCullagh and J.~A. Nelder.
\newblock \emph{Generalized linear models}, volume~37.
\newblock CRC press, 1989.

\bibitem[Menne et~al.(2012)Menne, Durre, Vose, Gleason, and Houston]{GHCN}
M.~J. Menne, I.~Durre, R.~S. Vose, B.~E. Gleason, and T.~G. Houston.
\newblock An overview of the global historical climatology network-daily
  database.
\newblock \emph{Journal of Atmospheric and Oceanic Technology}, 29\penalty0
  (7):\penalty0 897--910, 2012.

\bibitem[Naveau et~al.(2009)Naveau, Guillou, Cooley, and Diebolt]{naveau2009}
P.~Naveau, A.~Guillou, D.~Cooley, and J.~Diebolt.
\newblock Modelling pairwise dependence of maxima in space.
\newblock \emph{Biometrika}, 96\penalty0 (1):\penalty0 1--17, 2009.

\bibitem[Nelsen(2007)]{nelsen2007}
R.~B. Nelsen.
\newblock \emph{An introduction to copulas}.
\newblock Springer Science \& Business Media, 2007.

\bibitem[Nychka et~al.(2015)Nychka, Furrer, Paige, and Sain]{fieldsR}
D.~Nychka, R.~F. Furrer, J.~Paige, and S.~Sain.
\newblock fields: Tools for spatial data, 2015.
\newblock URL \url{www.image.ucar.edu/fields}.
\newblock R package version 8.10.

\bibitem[Pearl(2000)]{Pearl:2000}
J.~Pearl.
\newblock \emph{Causality - Models, Reasoning and Inference}.
\newblock Cambridge University Press, reprinted with corrections edition, 2000.

\bibitem[Quiroga et~al.(2002)Quiroga, Kreuz, and Grassberger]{quiroga2002event}
R.~Q. Quiroga, T.~Kreuz, and P.~Grassberger.
\newblock Event synchronization: a simple and fast method to measure
  synchronicity and time delay patterns.
\newblock \emph{Physical review E}, 66\penalty0 (4):\penalty0 041904, 2002.

\bibitem[{R Core Team}(2018)]{R}
{R Core Team}.
\newblock \emph{R: A Language and Environment for Statistical Computing}.
\newblock R Foundation for Statistical Computing, Vienna, Austria, 2018.
\newblock URL \url{http://www.R-project.org/}.

\bibitem[Rayner et~al.(2003)Rayner, Parker, Horton, Folland, Alexander, Rowell,
  Kent, and Kaplan]{HadISST2003}
N.~Rayner, D.~E. Parker, E.~Horton, C.~Folland, L.~Alexander, D.~Rowell,
  E.~Kent, and A.~Kaplan.
\newblock Global analyses of sea surface temperature, sea ice, and night marine
  air temperature since the late nineteenth century.
\newblock \emph{Journal of Geophysical Research: Atmospheres}, 108\penalty0
  (D14), 2003.

\bibitem[Ribatet and Sedki(2013)]{ribatet2013}
M.~Ribatet and M.~Sedki.
\newblock Extreme value copulas and max-stable processes.
\newblock \emph{Journal de la Soci{\'e}t{\'e} Fran{\c{c}}aise de Statistique},
  154\penalty0 (1):\penalty0 138--150, 2013.

\bibitem[Robbins(1955)]{robbins1955}
H.~Robbins.
\newblock An empirical bayes approach to statistics.
\newblock Technical report, COLUMBIA UNIVERSITY New York City United States,
  1955.

\bibitem[Runge(2014)]{RungeDiss}
J.~Runge.
\newblock \emph{Detecting and quantifying causality from time series of complex
  systems}.
\newblock PhD thesis, Humboldt-Universit{\"a}t zu Berlin,
  Mathematisch-Naturwissenschaftliche Fakult{\"a}t, 2014.

\bibitem[Schlather and Tawn(2003)]{schlather2003}
M.~Schlather and J.~A. Tawn.
\newblock A dependence measure for multivariate and spatial extreme values:
  Properties and inference.
\newblock \emph{Biometrika}, 90\penalty0 (1):\penalty0 139--156, 2003.

\bibitem[Segers(2012)]{segers2012}
J.~Segers.
\newblock Max-stable models for multivariate extremes.
\newblock \emph{REVSTAT-Statistical Journal}, 10\penalty0 (1):\penalty0 61--82,
  2012.

\bibitem[Smith(1990)]{smith1990}
R.~L. Smith.
\newblock Max-stable processes and spatial extremes.
\newblock 1990.

\bibitem[Spirtes et~al.(2000)Spirtes, Glymour, and Scheines]{SpGlSc:2000}
P.~Spirtes, C.~Glymour, and R.~Scheines.
\newblock \emph{Causation, Prediction, and Search}.
\newblock MIT Press, 2nd edition, 2000.

\bibitem[Steinhaeuser et~al.(2011)Steinhaeuser, Chawla, and
  Ganguly]{steinhaeuser2011}
K.~Steinhaeuser, N.~V. Chawla, and A.~R. Ganguly.
\newblock Complex networks as a unified framework for descriptive analysis and
  predictive modeling in climate science.
\newblock \emph{Statistical Analysis and Data Mining: The ASA Data Science
  Journal}, 4\penalty0 (5):\penalty0 497--511, 2011.

\bibitem[Stephenson(2002)]{stephenson2002}
A.~G. Stephenson.
\newblock evd: Extreme value distributions.
\newblock \emph{R News}, 2\penalty0 (2):\penalty0 31--32, 2002.

\bibitem[Thibaud et~al.(2016)Thibaud, Aalto, Cooley, Davison, Heikkinen,
  et~al.]{thibaud2016}
E.~Thibaud, J.~Aalto, D.~S. Cooley, A.~C. Davison, J.~Heikkinen, et~al.
\newblock Bayesian inference for the brown--resnick process, with an
  application to extreme low temperatures.
\newblock \emph{The Annals of Applied Statistics}, 10\penalty0 (4):\penalty0
  2303--2324, 2016.

\bibitem[Tibshirani(1996)]{tibshirani1996lasso}
R.~Tibshirani.
\newblock Regression shrinkage and selection via the lasso.
\newblock \emph{Journal of the Royal Statistical Society. Series B
  (Methodological)}, pages 267--288, 1996.

\bibitem[Trenberth et~al.(2018)Trenberth, Cheng, Jacobs, Zhang, and
  Fasullo]{trenberth2018}
K.~E. Trenberth, L.~Cheng, P.~Jacobs, Y.~Zhang, and J.~Fasullo.
\newblock Hurricane harvey links to ocean heat content and climate change
  adaptation.
\newblock \emph{Earth's Future}, 2018.

\bibitem[Tsonis and Roebber(2004)]{tsonis2004}
A.~A. Tsonis and P.~J. Roebber.
\newblock The architecture of the climate network.
\newblock \emph{Physica A: Statistical Mechanics and its Applications},
  333:\penalty0 497--504, 2004.

\bibitem[Tsonis et~al.(2006)Tsonis, Swanson, and Roebber]{tsonis2006}
A.~A. Tsonis, K.~L. Swanson, and P.~J. Roebber.
\newblock What do networks have to do with climate?
\newblock \emph{Bulletin of the American Meteorological Society}, 87\penalty0
  (5):\penalty0 585--595, 2006.

\bibitem[Tsonis et~al.(2008)Tsonis, Swanson, and Wang]{tsonis2008}
A.~A. Tsonis, K.~L. Swanson, and G.~Wang.
\newblock On the role of atmospheric teleconnections in climate.
\newblock \emph{Journal of Climate}, 21\penalty0 (12):\penalty0 2990--3001,
  2008.

\bibitem[Wadsworth et~al.(2017)Wadsworth, Tawn, Davison, and
  Elton]{wadsworth2017}
J.~Wadsworth, J.~A. Tawn, A.~Davison, and D.~M. Elton.
\newblock Modelling across extremal dependence classes.
\newblock \emph{Journal of the Royal Statistical Society: Series B (Statistical
  Methodology)}, 79\penalty0 (1):\penalty0 149--175, 2017.

\bibitem[Wahba(1990)]{wahba1990}
G.~Wahba.
\newblock \emph{Spline models for observational data}.
\newblock CBMS-NSF Regional Conference series in applied mathematics; 59.
  Society for Industrial and Applied Mathematics (SIAM, 3600 Market Street,
  Floor 6, Philadelphia, PA 19104), Philadelphia, PA., 1990.
\newblock ISBN 0898712440.

\bibitem[Wang et~al.(2016)Wang, Han, Stein, Kotamarthi, and Huang]{wang2016}
J.~Wang, Y.~Han, M.~L. Stein, V.~R. Kotamarthi, and W.~K. Huang.
\newblock Evaluation of dynamically downscaled extreme temperature using a
  spatially-aggregated generalized extreme value (gev) model.
\newblock \emph{Climate dynamics}, 47\penalty0 (9-10):\penalty0 2833--2849,
  2016.

\bibitem[Yamasaki et~al.(2009)Yamasaki, Gozolchiani, and
  Havlin]{yamasaki2009climate}
K.~Yamasaki, A.~Gozolchiani, and S.~Havlin.
\newblock Climate networks based on phase synchronization analysis track
  el-ni{\~n}o.
\newblock \emph{Progress of Theoretical Physics Supplement}, 179:\penalty0
  178--188, 2009.

\bibitem[Yan and Dey(2015)]{YanDey:book}
J.~Yan and D.~K. Dey.
\newblock \emph{Extreme Value Modeling and Risk Analysis}.
\newblock Chapman and Hall/CRC, New York, 1st edition, 2015.

\bibitem[Yuan and Lin(2007)]{yuan2007}
M.~Yuan and Y.~Lin.
\newblock Model selection and estimation in the gaussian graphical model.
\newblock \emph{Biometrika}, 94\penalty0 (1):\penalty0 19--35, 2007.

\bibitem[Zerenner et~al.(2014)Zerenner, Friederichs, Lehnertz, and
  Hense]{zerenner2014gaussian}
T.~Zerenner, P.~Friederichs, K.~Lehnertz, and A.~Hense.
\newblock A gaussian graphical model approach to climate networks.
\newblock \emph{Chaos: An Interdisciplinary Journal of Nonlinear Science},
  24\penalty0 (2):\penalty0 023103, 2014.

\end{thebibliography}

\end{document}